\documentclass{nsr}

\usepackage{amsmath,graphicx,array}
\usepackage{dcolumn,soul}%

\usepackage{amsthm}
\usepackage[version=4]{mhchem} 
\usepackage[figuresright]{rotating}%
\usepackage{algorithm, algorithmicx, algpseudocode}
\usepackage{listings}%
\usepackage{hyperref}

\makeatletter
\def\uns{\ifmmode\,\else$\,$\fi}%

\makeatother
 
\jvol{XX}
\jnum{X}
\jyear{Year}
\doi{10.1093/nsr/XXXX}
\received{XX XX Year}
\revised{XX XX Year}
\accepted{XX XX Year}

\begin{document}

\dhead{Review}

\subhead{EARTH  SCIENCE}

\title{Tipping Points and Cascading Transitions: Methods, Principles, and  Evidences}

\author{Sheng Fang$^{1}$}

\author{Ziyan Wang$^{1}$}

\author{J\"urgen Kurths$^{2,3,*}$}
\author{Jingfang Fan$^{1,2,*}$}

\affil{$^1$School of Systems Science, Beijing Normal University, 100875 Beijing, China}
\affil{$^2$Potsdam Institute for Climate Impact Research, 14412 Potsdam, Germany}
\affil{$^3$Department of Physics, Humboldt University Berlin, 10099 Berlin, Germany}

\authornote{\textbf{Corresponding authors.} Email: kurths@pik-potsdam.de, jingfang@bnu.edu.cn}

\abstract[ABSTRACT]{ 
This review synthesizes recent advancements in understanding tipping points and cascading transitions within the Earth system, framing them through the lens of nonlinear dynamics and complexity science. It outlines the fundamental concepts of tipping elements, large-scale subsystems like the Atlantic Meridional Overturning Circulation and the Amazon rainforest, and classifies tipping mechanisms into bifurcation-, noise-, and rate-induced types. The article critically evaluates methods for detecting early-warning signals, particularly those based on critical slowing down, while also acknowledging their limitations and the promise of non-conventional indicators. Furthermore, we explore the significant risk of cascading failures between interacting tipping elements, often modeled using conceptual network models. This shows that such interactions can substantially increase systemic risk under global warming. 
The review concludes by outlining key challenges related to data limitations and methodological robustness, and emphasizes the promising role of artificial intelligence and complex network science in advancing prediction and risk assessment of Earth system tipping dynamics.
}


\maketitle

\section{Introduction}

The ocean, covering about 71\% of Earth's surface, is a cornerstone of the Earth system, underpinning global climate regulation, sustaining marine biodiversity, and supporting socioeconomic development. 
It has absorbed nearly 93\% of the excess heat and about 30\% of anthropogenic \ce{CO2} emissions~\cite{bindoff2007observations,gruber2019oceanic}, harbors approximately 2.2 million species~\cite{mora2011many}, and contributes around USD~2.6~trillion annually to the global economy~\cite{oecd2025ocean}. 
However, under the combined pressures of global warming and human activities, the ocean is experiencing profound and systemic changes, including sea-level rise, warming, acidification, and deoxygenation~\cite{heinze2021quiet,collins2020extremes}. 
These interconnected processes are eroding marine ecosystem resilience and may push subsystems toward irreversible regime shifts.
This threat is especially acute in coastal regions owing to their position at the land–ocean boundary and heightened sensitivity to environmental changes. 
Since they are exposed simultaneously to terrestrial, marine, and human-induced disturbances, coastal systems often face amplified cumulative impacts that can accelerate ecological degradation and push them closer to critical thresholds~\cite{zhang2016regime}.
For example, global warming and anthropogenic nutrient inputs have caused severe deoxygenation in the Baltic Sea~\cite{carstensen2014deoxygenation}, which reduces fishery yields and promotes harmful algal blooms, which are extremely difficult to reverse through governance or intervention. 

The marine environment exemplifies a complex adaptive system, in which physical, biological, and socioeconomic components are dynamically coupled through nonlinear feedbacks operating across multiple spatial and temporal scales. 
Understanding such intertwined processes requires a holistic framework, as provided by Earth System Science (ESS). 
ESS conceptualizes the planet as an integrated system of interacting physical, chemical, biological, and anthropogenic subsystems~\cite{SteffenEmergence2020}. 
It seeks to reveal how natural processes and human activities jointly shape global dynamics, encompassing the climate system, biogeochemical cycles, and ecosystems. 
By integrating insights from geology, meteorology, oceanography, ecology, and the social sciences, ESS provides a comprehensive framework for addressing grand challenges in sustainability, resilience, and planetary boundaries, and for guiding the informed management of Earth's complex systems. 

Within the ESS framework, the concept of \textit{tipping points} has attracted increasing attention~\cite{LentonTipping2008,LentonClimate2019}. 
A tipping point refers to a critical threshold beyond which small perturbations can trigger large and often irreversible changes in a system's state. 
The corresponding large-scale subsystems of the Earth, which are known as \textit{tipping elements}, include the Atlantic Meridional Overturning Circulation (AMOC), the Amazon rainforest, the Greenland Ice Sheet, and the West Antarctic Ice Sheet, etc. Various methods, including early warning signals (EWS), have been designed to anticipate the tipping points~\cite{Scheffer2009Critical,surovyatkina2005fluctuation}. 
Importantly, these elements are not independent; their interactions can induce cascading effects that amplify systemic risk~\cite{liu2023teleconnections}. 
For instance, rapid Arctic sea-ice loss introduces freshwater into the North Atlantic, weakening the AMOC. 
This slowdown can reduce rainfall over South America, accelerating Amazon forest dieback. 
The resulting carbon release enhances global warming, potentially destabilizing the West Antarctic Ice Sheet~\cite{LentonClimate2019,WunderlingInteracting2021}. 
Recent assessments suggest that several major tipping elements are approaching, or may already have crossed, their critical thresholds~\cite{LentonClimate2019}, and that exceeding 1.5\,°C of global warming could trigger multiple, interconnected tipping events~\cite{Armstrong2022exceeding}. 

In this review, we summarize recent advances in understanding tipping points and cascading transitions within the Earth system. 
We first introduce the core concepts of tipping points and tipping elements, followed by theoretical foundations of tipping dynamics and methods for detecting early-warning signals. 
Next, we discuss conceptual and network-based models that capture cascading interactions among tipping elements. 
Finally, we highlight major challenges and emerging opportunities, emphasizing how artificial intelligence and complex network science can advance the prediction, understanding, and risk assessment of Earth system tipping dynamics.

\section{Tipping point and tipping element }
\label{sec:tipping_point}
As a basic form of  weak causality, the concept of ``tipping point'' originated from observations of abrupt sociological changes, capturing how small perturbations can lead to disproportionately large outcomes~\cite{Gladwell2006Tipping}. Over time, the term gained traction beyond the social sciences and has been increasingly adopted across diverse disciplines~\cite{Scheffer2009Critical}, including  ESS. A prominent application arises in climate science, where such thresholds are closely linked to what was previously termed \emph{abrupt climate change}~\cite{AlleyAbrupt2003}. 

Since the 1970s, the understanding of abrupt climate change has evolved from a theoretical notion to an empirically grounded paradigm. This shift was largely driven by high-resolution paleoclimatic archives, which revealed that Quaternary climate history was punctuated by rapid, high-magnitude oscillations, such as the Younger Dryas stadial~\cite{alley1993abrupt}. Broecker~\cite{broecker1987unpleasant} played a pioneering role in advancing the hypothesis that the climate system exhibits nonlinear behavior capable of rapid reorganization, particularly through perturbations to the AMOC. Subsequent syntheses of ice-core and marine sediment records further reinforced this view, demonstrating the recurrent nature of abrupt transitions throughout Earth's history~\cite{AlleyAbrupt2003}. 

Building upon this foundation, Lenton \emph{et al.}~\cite{LentonTipping2008} formally defined a tipping point as the critical threshold of a control parameter, $\rho_c$, beyond which an infinitesimal perturbation induces a significant change, $\hat{F}$, in the system state $F$ within a finite observation time $T$. This definition can be expressed as
\begin{equation}
    |F(\rho \ge \rho_c + \delta\rho | T) - F(\rho_c | T)| \ge \hat{F} > 0.
\end{equation}
They further introduced the concept of \textit{climate tipping elements}, referring to subcontinental-scale components of the Earth system that may cross a tipping point and transition into a qualitatively new state under the influence of climate change or anthropogenic forcing. These tipping elements are of particular policy relevance, as their activation could amplify global temperature rise and destabilize other parts of the system. Moreover, these elements are not independent, where transitions in one tipping element can trigger cascading effects across others, potentially leading to widespread and irreversible transformations of the Earth system. \par 

A typical example of a tipping element is the Amazon rainforest, which harbors over 10\% of Earth's terrestrial biodiversity and serves as one of the planet’s largest carbon sinks, absorbing vast amounts of atmospheric carbon dioxide through photosynthesis~\cite{SciencePanelAmazon2021,IPCC2021}.  
However, under the combined pressures of climate change and intensifying human activities, extensive regions of the Amazon rainforest are projected to undergo large-scale degradation and dieback in the coming decades~\cite{cano2022abrupt,parry2022evidence}.  
A recent study~\cite{flores2024critical} suggests that by 2050, approximately 10--47\% of the Amazon rainforest could be exposed to compounding disturbances, potentially triggering abrupt ecosystem transitions and amplifying regional climate change.  
Notably, the seminal paper in Ref.~\cite{LentonTipping2008} identified the Amazon rainforest as a potential tipping element of the Earth system, and its collapse could drive an additional global temperature rise of 3--4~°C, which underscores its pivotal role in regulating planetary stability. \par 

       \begin{table*}
         \centering  \begin{tabular}{|c|c|c|c|}
         \hline 
         Parameter & B-tipping& N-tipping & R-tipping\\
         \hline 
         $\mu$ & $ \frac{d \mu}{dt} = -4\times 10^{-4}$, $\mu(0) = 1.0$& 1.0 & 1.0\\
         \hline 
         $b_2(\times 10^{-5}{\rm K^{-2}})$ & 1.04& 1.690  & $(1-\lambda)b_{\rm ini} + \lambda b_{\rm fin}$ \\
         $a_2$ & 0.2& 0.6927 & $(1- \tilde{\rho}(\lambda) )a_{\rm ini} + \tilde{\rho}(\lambda)a_{\rm fin}$\\ 
         $\nu$ & 0& 1.0  & 0 \\
         \hline 
     \end{tabular}
     \caption{Parameter values for various type of tipping via Eq.~\eqref{eq:tipping_types}. 
     For B-tipping, the noise is absent with $\nu=0$ and the parameter $\mu$ is decreased from 1 at a rate of $4 \times 10^{-4}$ per year. For N-tipping, the parameter is fixed to $1.0$ and the amplitude $\nu=1$.  
     For R-tipping, the initial and final values of parameters $b_2$ are fixed as $b_{\rm ini} = 1.69 \times 10^{-5}$ and $b_{\rm fin} = 1.835 \times 10^{-5}$. The parameter $\lambda$ controls the ariability of $b_2$, which is the solution of equation $d\lambda/dt = \rho \lambda(1-\lambda)$. In the simulation, the initial parameters of $\lambda$ is set to $10^{-6}$. 
     Similar case is applied for $a_2$ with $a_{\rm ini} = 0.6927$ and $a_{\rm fin} = 0.8168$, and $\tilde{\rho}(\lambda)$ is the controling parameter such that $(c_1 \mu b_2)^2 -4c_1\mu a_2$ is constant at a rate proportional to $\rho$.}
     \label{tab:tipping_types_parameter}
 \end{table*}

 \begin{figure*}[t]
     \centering      
     \includegraphics[width=0.9\linewidth]{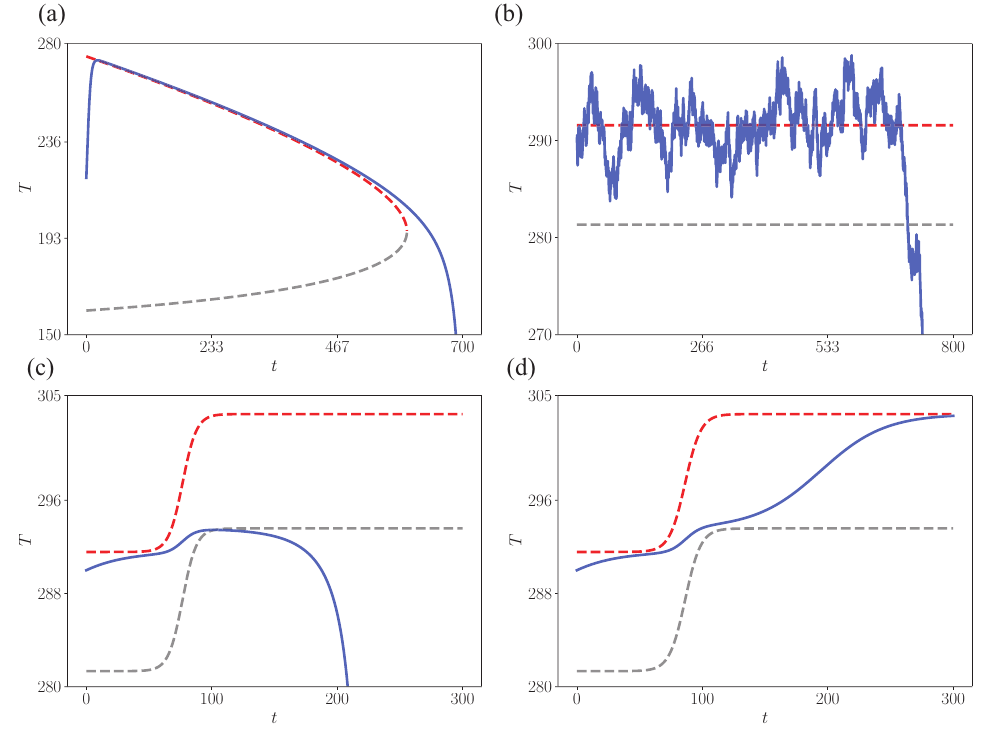}
     \caption{ Illustrations of trajectories for the Sutera-Fraedrich model Eq.~\eqref{eq:tipping_types}~\cite{Fraedrich1979Catastrophes, Sutera1981On} showing the presence of three various types of tipping with  (a) B-tipping, (b) N-tipping, (c)-(d) R-tipping. This figure is a replot of Figure 7 in Ref.~\cite{Ashwin2012Tipping}, and the controlled parameters are summarized in Table~\ref{tab:tipping_types_parameter}. 
     The horizontal axis is time with dimension yr, and the vertical axis is the temperature with dimension K. 
     The solid lines show system trajectories while the dashed lines show the location of the fixed point, where red dashed denotes the stable fixed point and the gray dashed line is for the unstable fixed point. 
     For R-tipping, the parameter $\rho$ takes $0.16$ for (c) and $0.18$ for (d), and the critical rate is approximately 0.175. }
     \label{fig:tipping_types}
 \end{figure*}
\section{Bifurcation and tipping types}
To model the tipping process, we consider a generalized dynamical system,
\begin{equation}
    \frac{dx}{dt} = F(x, \rho(t)),
    \label{eq:dynamic_system}
\end{equation}
where $x$ represents the system state variable and $\rho(t)$ is a time-dependent control parameter.  
In the special case where $\rho$ is constant, Eq.~\eqref{eq:dynamic_system} reduces to an autonomous system that may evolve toward a quasi-static attractor.  
Based on the underlying mechanisms,  whether state transitions are triggered by internal dynamics or external perturbations, tipping phenomena can generally be classified into three causally distinct types: ({\romannumeral 1}). \textit{bifurcation-induced tipping} (B-tipping), ({\romannumeral 2}). \textit{noise-induced tipping} (N-tipping), and ({\romannumeral 3}). \textit{rate-dependent tipping} (R-tipping)~\cite{Ashwin2012Tipping,lenton2023global}.

({\romannumeral 1}). B-tipping occurs when a slow, continuous variation in system parameters (e.g., external forcing or climate trends) drives the system through a bifurcation, causing a stable state, such as a fixed point or a periodic orbit, to lose stability or vanish. This loss of stability leads to an abrupt transition to an alternative attractor.  
({\romannumeral 2}). In contrast, N-tipping arises in multistable systems, where stochastic fluctuations (e.g., interannual variability in the climate system) can push the system across a potential barrier, inducing a noise-driven transition from one stable state to another.  
({\romannumeral 3}). R-tipping occurs when the rate of parameter change is sufficiently rapid  such that the system cannot adiabatically follow its quasi-static equilibrium. As a result, the system departs from the expected attractor before the bifurcation point is reached, or it may delay the transition,  which leads to an early or late tipping event.  

\par

To illustrate these concepts, we employ the ``zero-dimensional'' global energy balance model~\cite{Fraedrich1979Catastrophes, Sutera1981On,lucarini2019transitions}, expressed as
\begin{align}
    dT &= f(T)\, dt + \sqrt{\nu}\, dW, \nonumber \\
    f(T) &= c_0[-T^4 + c_1\mu(b_2T^2 - a_2)].
    \label{eq:tipping_types}
\end{align}
This model describes the temporal evolution of the global mean surface temperature $T$, representing the balance between incoming solar radiation and outgoing longwave radiation.  
The term $dW$ denotes a normalized Wiener process (white noise) with amplitude $\sqrt{\nu}$, capturing stochastic perturbations from the environment.  
All parameters are scaled such that $f(T)$ has the dimension of ${\rm K\,yr^{-1}}$.  
Under these assumptions, the equilibrium-state parameters for $\nu=0$ can be computed as $c_0 = 1.10869\times 10^{-8}$${\rm K^{-3}\,yr^{-1}}$ and $c_1 = 9.7137 \times 10^9$${\rm K^4}$.  

The parameters $\mu$, $b_2$, and $a_2$ are treated as control variables.  
Table~\ref{tab:tipping_types_parameter} summarizes the parameter settings used to simulate different types of tipping, while the corresponding time evolutions are shown in Fig.~\ref{fig:tipping_types}.  
For the B-tipping case, the system evolves toward a stable fixed point when far from the critical threshold. As the tipping point is approached, the distance between the stable and unstable equilibria decreases, causing the system's trajectory to deviate from the stable branch.  
In the N-tipping scenario, the system exhibits bistability in the absence of noise. Before the transition occurs, the state fluctuates around the stable fixed point, but may stochastically jump to the alternative attractor (e.g., near $T \approx -300$ K) once noise is introduced.  
R-tipping differs fundamentally from the previous two types: the system initially follows a quasi-steady trajectory between the stable and unstable branches~\cite{Ashwin2012Tipping}. When the rate of parameter variation $\rho(t)$ exceeds a critical threshold, the system departs from the stable branch before reaching the bifurcation point, triggering a rapid transition, as illustrated in Fig.~\ref{fig:tipping_types}(c).  \par

 \subsection{Early warning signals and tipping point detection}

Since tipping events can cause profound and often irreversible shifts in system states, detecting and anticipating their onset is of paramount importance. Yet, the intrinsic complexity of empirical systems often makes it difficult to determine the critical threshold in advance.
Fortunately, many complex systems exhibit  \emph{universal}  signatures, such as rising variance, autocorrelation, and skewness, as they approach critical transitions. These phenomena are collectively referred to as \emph{early-warning signals} (EWS)~\cite{Scheffer2009Critical}. 
A range of EWS detection methods have been developed and are summarized in Table~\ref{tab:ews}, following Ref.~\cite{DakosTipping2024}. 
Among these approaches, one of the most widely applied is based on the concept of \emph{critical slowing down} (CSD)~\cite{SchefferEarlywarning2009}, which describes the progressive decline in a system’s recovery rate as it nears a bifurcation point in B-tipping scenarios. 
Indicators derived from CSD, such as variance, temporal and spatial correlations, and recovery rate, are particularly useful for detecting imminent transitions, as summarized in Table~\ref{tab:ews}.  
\par

To illustrate the underlying mechanism, consider a stochastic saddle-node bifurcation system governed by
\begin{equation}
    \frac{dx}{dt} = a x^3 + bx + \phi(\rho) + \xi(t),
    \label{eq:Btipping}
\end{equation}
where $\phi(\rho)$ represents a control parameter, and $\xi(t)$ denotes Gaussian white noise. 
In the absence of noise, the stability analysis yields the critical thresholds $\phi_c^{\pm}=  \pm \sqrt{-4a^3b^3 / 27a^4}$, under the intrinsic condition $a<0$ and $b>0$. 
When $\phi < \phi_c^-$ or $\phi > \phi_c^+$, the system possesses a single equilibrium; for $\phi_c^{-} < \phi < \phi_c^+$, two stable equilibria coexist with one unstable state. 
As $\phi$ approaches $\phi_c^+$, all equilibria merge and vanish, marking a classical saddle-node bifurcation. 
\par

Let us consider a specific case with $a=-1$ and $b=1$, for which $\phi_c = \pm \frac{2\sqrt{3}}{9}$. 
Near $\phi_c^-$, the equilibrium position is $x_c = \frac{\sqrt{3}}{3}$. 
Introducing a small perturbation $\epsilon$ around this equilibrium, $x = x_c + \epsilon$, a local linearization gives
\begin{align}
    \frac{d x_c}{dt} +  \frac{d\epsilon }{dt} &= f(x, \phi), \\
    f(x, \phi) &\approx f(x_c, \phi) + \left.\frac{\partial f(x,\phi)}{\partial x}\right|_{x_c} \epsilon + O(\epsilon^2).
\end{align}
For $x \to x_c$, the recovery rate is $\lambda(\phi) = \left.\frac{\partial f(x,\phi)}{\partial x}\right|_{x_c} = -3x_c^2 + 1$. 
At the critical point, $\lambda(\phi) \to 0$, implying that small perturbations decay increasingly slowly—a hallmark of CSD. \par

This critical slowing down manifests as increasing temporal and spatial correlations. 
For a discrete time series, the lag-1 autocorrelation (AC1) is defined as
\begin{equation}
    {\rm AC1} = \frac{\sum_{t=2}^{T} (x_{t} -\bar{x})(x_{t-1} -\bar{x})}{\sum_{t=1}^{T} (x_t-\bar{x})^2}.
\end{equation}
Assuming small perturbations ($\bar{x} \approx x_c$), we have ${\rm AC1} \approx e^{\lambda}$. 
As the system approaches the threshold, $\lambda \to 0$ and hence ${\rm AC1} \to 1$. 
Similarly, for perturbations obeying 
\begin{equation}
    x_{t+1} -\bar{x} = e^{\lambda} (x_{t}- \bar{x}) + \sigma \xi(t),
\end{equation}
where $\sigma$ denotes the amplitude of the white noise, the variance satisfies
\begin{equation}
    {\rm Var}(x_{t+1} - \bar{x}) = e^{\lambda} {\rm Var}(x_t  - \bar{x}) + \sigma.
\end{equation}
Thus, ${\rm Var} = \sigma / (1 - e^{\lambda})$, which diverges as $\lambda \to 0$, signaling proximity to the critical point. 
\par

Another commonly used technique is \emph{detrended fluctuation analysis} (DFA)~\cite{held2004detection}, designed to quantify long-term correlations and scaling properties in noisy, nonstationary time series. 
DFA removes local trends to estimate a scaling exponent $\alpha$, which approaches unity near a tipping point. 
Unlike AC1, which captures short-term (lag-1) correlations, DFA reveals long-range memory effects and helps mitigate false positives arising from slow drifts or nonstationarity~\cite{peng1994mosaic,livina2007modified}. 
However, it generally requires longer time series and is computationally more demanding. 
\par

\begin{figure}
     \centering
     \includegraphics[width=1.0\linewidth]{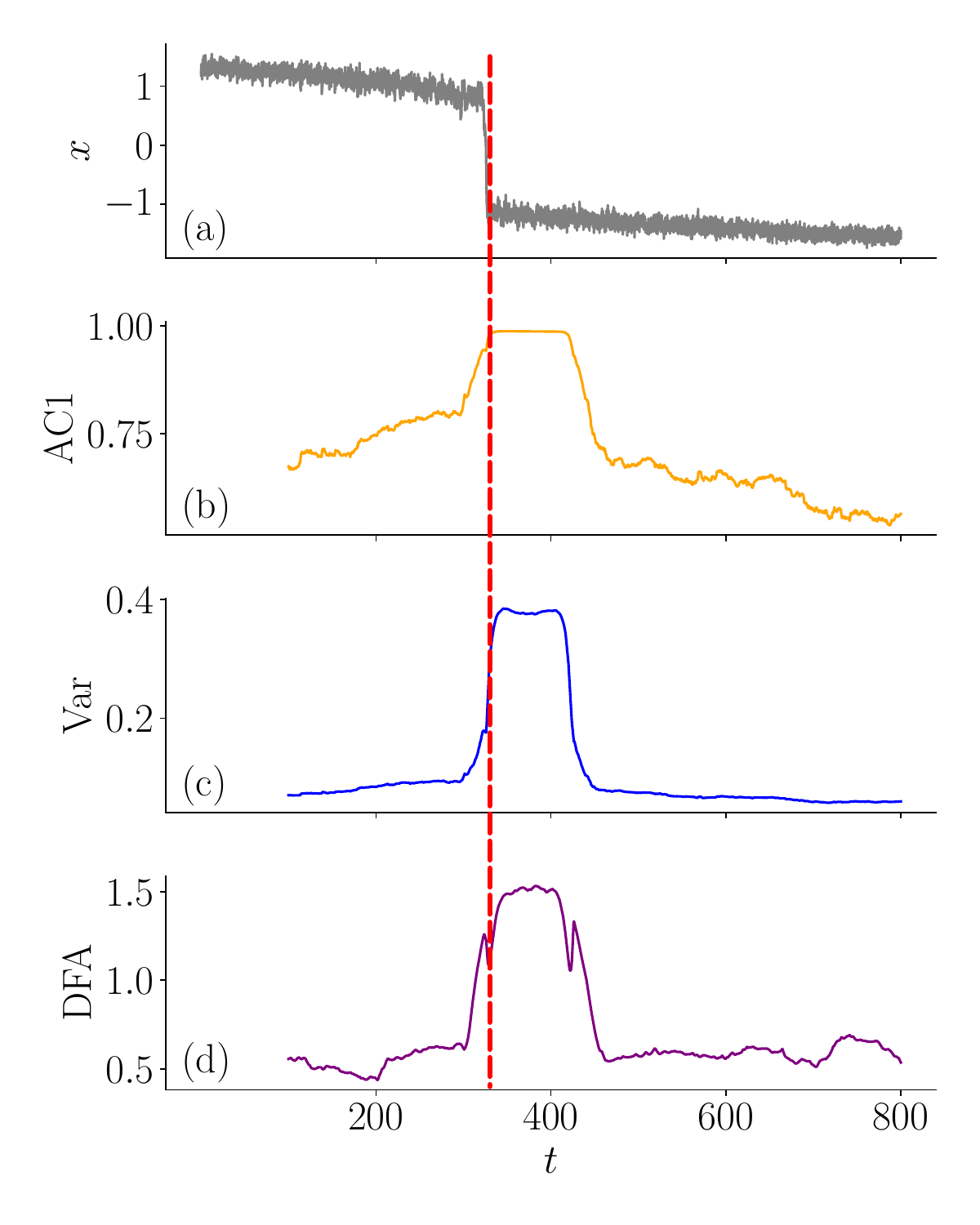}
     \caption{Illustration of the EWS of the tipping dynamic. The B-tipping dynamics is governed by Eq.~\eqref{eq:Btipping} (a) with three types of EWS indicators (b) AC1, (b) Var, and (d) DFA.}
     \label{fig:Btipping}
 \end{figure}

In Fig.~\ref{fig:Btipping}, we illustrate the dynamics of Eq.~\eqref{eq:Btipping} and its corresponding CSD indicators. 
The control parameter is set as $\phi = -1 + 0.004t$~\cite{BoersDestabilization2025}.
As the system approaches the bifurcation, both AC1 and variance increase markedly, with AC1 reaching a plateau and subsequently declining after the transition has occurred. 
\par

Even though CSD-based EWS have been successfully applied across a wide range of disciplines, they still face notable limitations. 
These methods are fundamentally grounded in the assumption that perturbations can be linearized near an equilibrium state, an assumption that holds primarily for B-tipping scenarios. 
However, this premise breaks down in the cases of N-tipping and R-tipping. 
In N-tipping, stochastic fluctuations exert strong, nonlinear influences that cannot be captured by linearization. 
In R-tipping, the system trajectory departs from any quasi-static equilibrium due to the rapid variation of external parameters, rendering linearization-based approaches invalid. 
To address these shortcomings, several non-CSD-based EWS indicators have been proposed. 
For example, skewness~\cite{GuttalChanging2008}, which quantifies the asymmetry of the system-state distribution, can reveal nonlinear shifts toward a tipping point by identifying increasing occurrences of extreme deviations in one direction. 
Another notable indicator is flickering~\cite{WangFlickering2012}, a phenomenon in which the system intermittently switches between alternative stable states under stochastic forcing. 
Such behavior reflects the progressive loss of resilience as the system begins to explore adjacent basins of attraction. 
Together, these non-CSD-based indicators provide valuable diagnostic tools for anticipating critical transitions in systems where traditional CSD-based methods fail to apply.
\par

EWS approaches have been widely implemented in the analysis of diverse climate subsystems~\cite{boers2021observation,boulton2022pronounced,ben2023uncertainties,liu2023teleconnections}. 
For instance, the authors presented empirical evidence suggesting that the AMOC may be approaching a tipping point~\cite{boers2021observation}. 
Using a century-long dataset (1900–2020) of sea surface temperature and salinity from the subpolar North Atlantic, which is a region particularly sensitive to AMOC variability, they applied statistical EWS indicators to assess long-term stability. 
Their analysis revealed a pronounced increase in both autocorrelation and variance over time, hallmarks of CSD that indicate a loss of dynamical resilience. 
These results suggest that the AMOC has been progressively weakening, particularly since the mid-20th century, and may be nearing a tipping point. 
While the precise timing of a potential collapse remains uncertain, this study provides a robust empirical warning that such an event could be imminent, underscoring the urgent need for enhanced observation and proactive climate mitigation efforts.
\par

\begin{table*}[h!]
\centering
\caption{Summary of methods to detect  EWS for tipping points. The table originates from Ref.~\cite{DakosTipping2024}.}
\label{tab:ews}
\scalebox{0.8}{ 
\begin{tabular}{|c|c|}
\hline
\textbf{CSD-based} & \textbf{Non-CSD-based } \\
\hline
  Variance (temporal/spatial) & Skewness (temporal/spatial) \\
  Autocorrelation (temporal/spatial) &  Kurtosis (temporal) \\
  Return rate and/or time (temporal) & Potential analysis (temporal) \\
  Detrended fluctuation analysis (temporal)  & Turing patterns (spatial) \\
  Spectral reddening (temporal)   & Hurst exponent (spatial/temporal) \\ 
  Variance–covariance eigenvalue (temporal) & Fisher information (temporal) \\
  Dynamic eigenvalue (temporal) & Mean exit time Fokker–Planck (temporal) \\
  repair time (spatial)    & Nonlinearity (temporal) \\
  Discrete Fourier Transform (spatial)   & Trait statistical changes (temporal) \\
  Generalized models (temporal)  & Machine learning approach (temporal) \\
  Time-varying AR(p) models (temporal)   & Average flux (temporal) \\
  Probabilistic time-varying AR(p) (temporal) & Quickest detection method (temporal) \\
  Machine learning approach (temporal) recovery length (spatial) &  Conditional heteroscedasticity  (temporal/spatial) \\ 
  Speed of travelling waves (spatial)    & Patch size distributions (spatial) \\
  & Kolmogorov complexity (spatial) \\
  & Network properties (spatial/temporal) \\
  & Drift–diffusion–jump models (temporal) \\
  & Threshold AR($p$) models (temporal) \\
  & Likelihood ratio (temporal) \\
\hline
\end{tabular} }
\end{table*}

\section{Cascading transitions} 

\begin{figure*}
    \centering
    \includegraphics[width=1.0\linewidth]{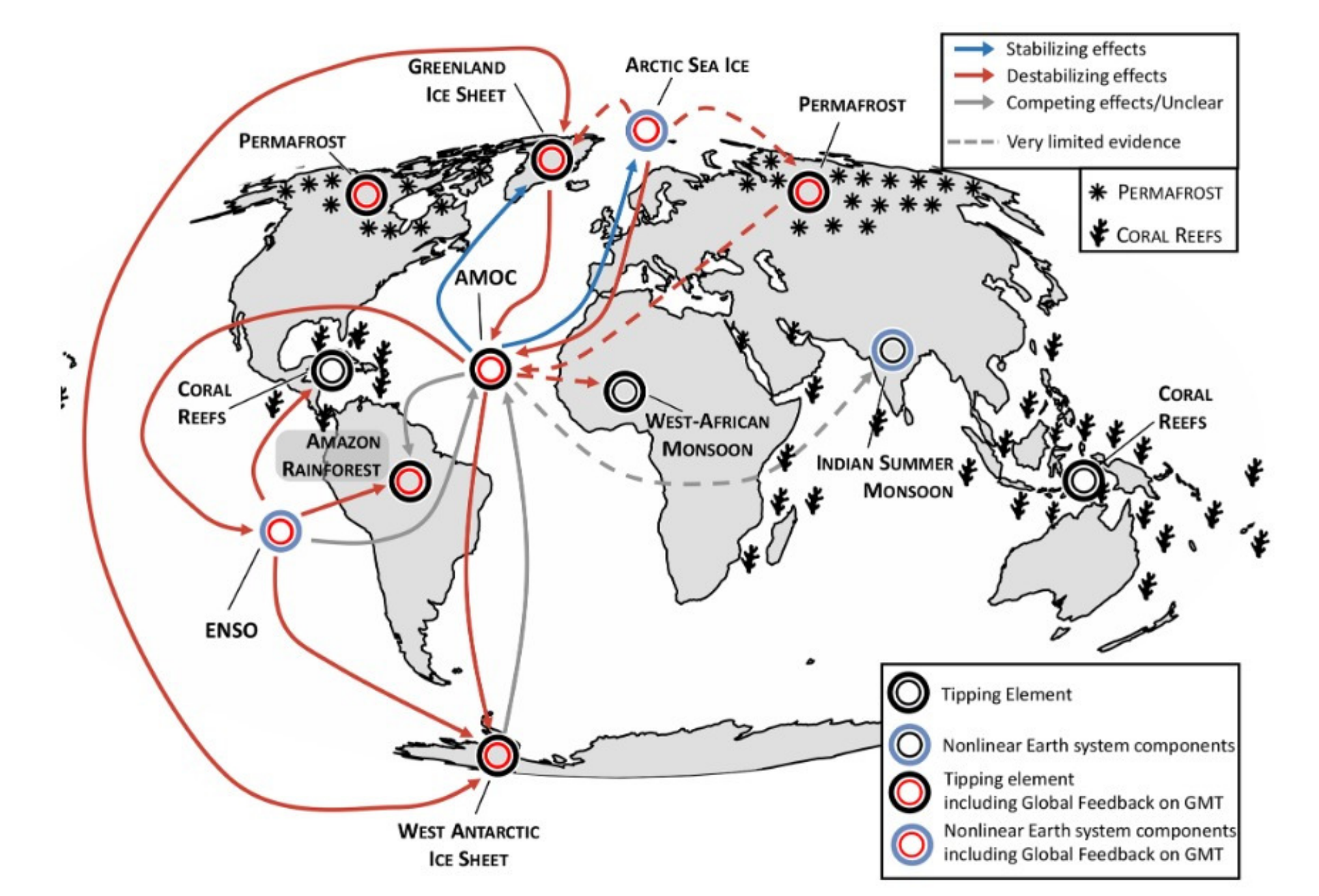}
    \caption{Illustration of interactions between tipping elements on a world map, and this figure originates from Ref.~\cite{DakosTipping2024}. All tipping elements discussed in this review article are shown together with their potential connections. The causal interactions links can have stabilizing (blue), destabilizing (red), or unclear (gray) effects. Tipping elements that exert a notable feedback on global mean temperature when they tip are denoted by a red inner ring. This temperature feedback can be positive (i.e., amplifying warming, as likely for the permafrost, the Arctic sea ice, the Greenland and West Antarctic ice sheets, the Amazon rainforest, and ENSO) or negative (i.e., dampening warming, as likely for the AMOC).}
    \label{fig:cascade}
\end{figure*}

From a risk-assessment perspective, the catastrophic impact of crossing a tipping point arises not only from the transition of an individual subsystem, but also from its capacity to trigger cascading effects across other tipping elements through complex interactions. 
Such cascades can propagate through interconnected components of the Earth system, including the cryosphere, biosphere, and ocean circulation, forming feedback loops that amplify the initial perturbation and may lead to large-scale, potentially irreversible consequences. 
For instance, melting of the Greenland Ice Sheet can disrupt the AMOC~\cite{rahmstorf2015exceptional,caesar2018observed}, which in turn may alter monsoon patterns and accelerate Arctic permafrost thaw~\cite{DekkerCascading2018}, as shown in Fig.~\ref{fig:cascade}.
A recent study indicates that exceeding a global warming threshold of $1.5^\circ{\rm C}$ could trigger multiple climate tipping points~\cite{Armstrong2022exceeding}, and that some critical thresholds might even lie below this level~\cite{LentonTipping2008}. 
This highlights that the cumulative risk of interacting tipping elements is substantially greater than that inferred from analyzing them in isolation, especially when considering synergistic effects and domino-like cascades of tipping events.
\par

Although the terms ``cascade'' and ``domino effect'' are often used interchangeably across disciplines, a key distinction lies in whether causality is a necessary condition for a cascade~\cite{lenton2020tipping,rocha2018cascading,brummitt2015coupled,dekker2018cascading}. 
Klose \emph{et al.}~\cite{KloseWhat2021} proposed three distinct patterns of multiple tipping dynamics: the \emph{domino cascade}, the \emph{two-phase cascade}, and the \emph{joint cascade}. 
To illustrate these, we consider a leading system $X_1$ and a following system $X_2$, both driven by an external control parameter $c_2$. 
As $c_2$ changes, the leading system $X_1$ may tip and, through a direct coupling, induce tipping in $X_2$, a process known as a \emph{domino cascade}. 
This form emphasizes causality, where tipping signals propagate from the leading to the following system through explicit interaction, and this has been widely used to model signal transmission between tipping elements in the Earth system~\cite{wunderling2020basin,WunderlingInteracting2021}. 
In contrast, a \emph{two-phase cascade} describes a situation in which both systems $X_1$ and $X_2$ cross their tipping points sequentially as $c_2$ varies, yet the tipping of $X_1$ does not immediately cause that of $X_2$; rather, further sustained change in $c_2$ leads to the subsequent tipping of $X_2$. 
Thus, causality is not a necessary condition for this type of cascade. 
This mechanism has been invoked to explain the consecutive tipping of the AMOC (as $X_1$) and the Antarctic Ice Sheet (as $X_2$) during the Eocene–Oligocene transition, accounting for the stepwise shifts observed in oxygen isotope records~\cite{coxall2005rapid,tigchelaar2011new,dekker2018cascading}. 
In a \emph{joint cascade}, two systems $X_1$ and $X_2$ tip simultaneously, as observed in some spatially extended bistable ecosystems~\cite{van2005implications,dakos2010spatial}.
\par

Compared with single-element tipping processes, compound tipping involving cascade effects poses far greater challenges for scientific analysis. 
These challenges span the identification, diagnosis, early warning, and prediction of tipping events, as well as the mathematical modeling and quantitative characterization of cascade dynamics themselves. 
In Ref.~\cite{DekkerCascading2018}, the authors developed a dynamical-systems framework to describe deterministic cascading processes between coupled subsystems. 
The conceptual model can be formulated as
\begin{equation}
\begin{cases}
    \dfrac{dx}{dt} = F_{X_1}(x, \rho), \\
    \dfrac{dy}{dt} = F_{X_2}(y, x),
    \label{eq:cascade_autonomous}
\end{cases}
\end{equation}
where $x$ and $y$ represent the state variables of the leading system $X_1$ and the following system $X_2$, respectively. 
The functions $F_{X_1}$ and $F_{X_2}$ characterize the underlying bifurcation types, such as fold or Hopf bifurcations. 
The leading system $X_1$ is driven by an external control parameter $\rho$, while the following system $X_2$ is influenced by the state of $X_1$ through linear coupling. 
Through numerical simulations, the study demonstrated that traditional EWS indicators may fail for the following system: tipping signals transferred from $X_1$ to $X_2$ can produce apparent increases in autocorrelation and variance even when $X_2$ remains far from its own tipping threshold.
\par

To capture more pairwise interactions and incorporate the influence of global warming, where the external driver, such as the global mean surface temperature $T$, varies with time $t$. Wunderling \emph{et al.}~\cite{WunderlingInteracting2021} proposed a stylized dynamic network model to describe cascading effects among climate tipping elements. 
In contrast to the deterministic and autonomous framework in Eq.~\eqref{eq:cascade_autonomous}, this model is nonautonomous. 
Moreover, it extends beyond the interaction between two tipping elements to formulate a dynamic network framework that encompasses multiple elements under temporally varying external forcing. 
The model assumes a double-fold bifurcation with a linear coupling term~\cite{brummitt2015coupled,klose2020emergence,kronke2020dynamics}:
\begin{align}
        \label{eq:cascade_network_model}
        \frac{d x_i}{dt} = \Bigg[ 
        & -x_i^3 + x_i + \sqrt{\frac{4}{27}} \cdot \frac{\Delta {\rm GMT}(t)}{T_{{\rm cri},i}} \nonumber \\
        & + d \cdot \sum_{j \neq i} \frac{s_{ij}}{10} (x_j + 1) 
        \Bigg] \frac{1}{\tau_i}.
\end{align}
Here, the first three terms represent the intrinsic bifurcation dynamics of a single tipping element $x_i$, while the last term captures the interactions from other elements $j$. 
The parameter $T_{{\rm cri},i}$ denotes the estimated critical threshold of tipping element $i$, and $\Delta {\rm GMT}(t)$ is the increase in global mean surface temperature relative to pre-industrial levels, which has been generalized to include the  GMT feedback induced by the crossing of a tipping point~\cite{bdolach2025tipping}. 
In the coupling term, $d$ represents the overall interaction strength, and $s_{ij}$ quantifies the connection strength between tipping elements.
The sign of $s_{ij}$ denotes positive or negative effects among various tipping elements, as illustrated in Fig.~\ref{fig:cascade}. 
For example, freshwater flux and sea-level rise act as links from the Greenland Ice Sheet to the AMOC and West Antarctic Ice Sheet, respectively~\cite{kriegler2009imprecise}. 
Compared with the fold-bifurcation model in Eq.~\eqref{eq:Btipping}, an additional parameter $\tau_i$ is introduced to represent the characteristic timescale of tipping. 
Parameter values for $T_{{\rm cri},i}$, $s_{ij}$, and $\tau_i$ can be obtained from existing studies~\cite{Armstrong2022exceeding}, while $d$ is typically varied within the range $[0,1]$ to account for uncertainty in real-world coupling strength. 
In the absence of coupling, the system approaches its bifurcation threshold as $\Delta {\rm GMT}$ rises toward $T_{{\rm cri},i}$. 
When coupling is introduced, the timing of tipping events can either advance or be delayed, depending on the net effect of positive and negative links. 
Through large ensembles of Monte Carlo simulations propagating parameter uncertainties, the authors investigated cascading tipping among four key elements, the Greenland and West Antarctic Ice Sheets, the AMOC, and the Amazon rainforest. 
Their findings reveal that interactions tend to destabilize the network, with the polar ice sheets often acting as initiators of cascading transitions. 
This aligns with evidence suggesting that several cryospheric tipping elements are already approaching critical thresholds~\cite{portner2019ocean,LentonClimate2019}.
\par

The network model in Eq.~\eqref{eq:cascade_network_model} was further extended to assess tipping risks under global temperature overshoot scenarios~\cite{WunderlingGlobal2023}, in which temporary exceedance of the $1.5^\circ{\rm C}$ or $2^\circ{\rm C}$ targets, the so-called ``overshoot'', substantially elevates the probability of irreversible cascading events. 
Using the stylized network model, the authors demonstrated that triggering one tipping element (e.g., Greenland Ice Sheet collapse) during an overshoot phase can ``lock in'' that change, subsequently destabilizing other elements (e.g., the AMOC) by reducing their tipping thresholds. 
This process greatly increases the likelihood of a domino-like chain of climate system failures. 
Crucially, their analysis shows that temperature overshoots make damaging cascades roughly three times more likely than pathways that stabilize temperatures without an overshoot. 
They further estimate that temporary overshoots can increase tipping risks by up to 72\% compared with non-overshoot scenarios, even if the long-term equilibrium temperature remains within the Paris Agreement range. 
These results underscore that avoiding even temporary exceedance of critical warming limits is essential to prevent self-reinforcing and potentially irreversible climate tipping cascades.
\par

\section{Challenges}

In this review, we have reviewed fundamental concepts and recent advances in the study of tipping points and cascading effects from the perspective of dynamical systems and complexity science, with particular emphasis on EWS and their detection. 
A stylized network-based dynamic model was introduced to describe interactions and cascading behaviors among multiple climate tipping elements. 
Nevertheless, research in this field still faces substantial challenges, which can be broadly categorized into two aspects: \emph{data} and \emph{methodology}.
\par

\subsection{1. Data quantity}
Thus far, most studies remain largely theoretical rather than data-driven, primarily due to the scarcity and heterogeneity of high-quality observational datasets suitable for diagnosing tipping points.  
A major challenge lies in the stringent data requirements of EWS methods, which typically demand long, high-resolution, and continuous time series.  
In practice, however, many climate records are either too short or contain missing values, resulting in discontinuities that severely hinder reliable EWS estimation.  
This limitation is particularly pronounced in the context of ocean observations, which are inherently sparse, spatially biased, and subject to substantial measurement uncertainties~\cite{elipot2022overcoming,abraham2013review,kennedy2014review}.  
A key limitation arises from the uneven spatial distribution of oceanic observations across the globe, with vast data gaps in remote and deep-sea regions.  
In particular, the Argo program~\cite{jayne2017argo}, consisting of over 3,000 autonomous profiling floats that continuously measure temperature and salinity, represents a cornerstone of the modern ocean observing system and complements satellite-based missions.  
Nevertheless, significant challenges persist: observational coverage remains biased toward the Northern Hemisphere due to logistical and funding constraints, while float operations are limited in the Southern Ocean and polar regions because of technical factors, such as the lack of effective ice-avoidance mechanisms.  
These factors collectively lead to pronounced spatial heterogeneity and substantial observational gaps~\cite{abraham2013review}.  
Moreover, direct observations of the deep ocean are intrinsically challenging.  
Extreme pressure, perpetual darkness, and inaccessibility make it exceptionally difficult to obtain high-precision, high-resolution in-situ measurements in abyssal environments.  
Consequently, available datasets are often biased or error-prone but are nonetheless used to initialize and validate large-scale climate models.  
This dependence inevitably compromises the reliability of model simulations and amplifies uncertainties in future climate projections, particularly in processes such as deep and bottom water formation, which remain poorly constrained in most models~\cite{heuze2020antarctic}. \par

In addition, all measurement systems are subject to instrumental errors, sensor drift, and calibration biases that can introduce systematic inaccuracies. 
Uncertainties also arise during the data processing and integration stages. 
For instance, gridding sparse point measurements or deriving geophysical quantities from satellite signals inevitably involves interpolation assumptions and empirical models that propagate error. 
Integrating heterogeneous datasets, from satellites, in situ measurements, and reanalysis products, further requires rigorous cross-calibration and homogenization to ensure consistency across spatial and temporal scales~\cite{abraham2013review,smith2022reliability}. 
Moreover, uncertainties in marine ecosystem projections arise from multiple sources, including structural (model) uncertainty, initialization and internal variability, parameter uncertainty, and scenario dependence~\cite{kennedy2014review,payne2016uncertainties}.
\par

\subsection{2. Methods of EWS and cascade}
Despite their theoretical appeal, EWS methodologies face substantial practical and conceptual limitations that constrain their predictive reliability~\cite{dakos2015resilience,burthe2016early,boettiger2013early,dai2015relation,DakosTipping2024}. 
Dakos \emph{et al.}~\cite{dakos2015resilience} noted that not all regime shifts correspond to true tipping points, meaning that classical CSD-based indicators may fail to manifest even when abrupt transitions occur. 
A recent meta-analysis~\cite{DakosTipping2024} reviewing over 200 studies found mixed results, with many reporting poor or negative predictive performance, especially in ecological contexts. 
Notably, nearly all negative cases relied solely on CSD-based early warnings. 
In a large scale empirical test, Burthe \emph{et al.}~\cite{burthe2016early} analyzed long term abundance time series from 55 taxa (126 datasets) across multiple trophic levels in marine and freshwater ecosystems. 
Their findings revealed that true positives were rare, only 9\% (16 of 170) for variance and 13\% (19 of 152) for autocorrelation, while false positives occurred more frequently than false negatives (53\% vs. 38\% for variance; 47\% vs. 40\% for autocorrelation). 
False detections were observed across all decades and ecosystem types. 
Additionally, most EWS techniques are developed for single external drivers, and their fundamental assumptions can fail when multiple drivers co-vary or interact~\cite{dai2015relation}.
\par

A further challenge in studying cascading tipping phenomena lies in the parameterization of stylized network models~\cite{WunderlingInteracting2021}.  
In most existing frameworks, the estimation of critical thresholds and coupling strengths relies heavily on expert judgment rather than data-driven inference, which inevitably introduces subjectivity and constrains the generalizability of these models.  
At present, methods for empirically constraining such parameters from observed or simulated cascade events remain largely underdeveloped.  
Further complexity arises from multi-timescale interactions among different climatic tipping elements, which challenge both the formulation and calibration of network-based dynamical models.  
This methodological gap represents a fundamental limitation: without objective, data-informed calibration, stylized network models possess limited predictive capability and reduced practical utility for assessing systemic risks or designing effective resilience and mitigation strategies.

\par

\section{Opportunities}
Given these challenges, new methodological paradigms are urgently needed. 
In particular, the integration of artificial intelligence (AI) and complex network science~\cite{barabasi2013network} offers a promising avenue for advancing the study of tipping points and cascading transitions.
\par

\subsection{1. Artificial intelligence}

Recent advances in AI have rapidly accelerated its application in Earth system science, leading to the emergence of an interdisciplinary AI–Earth Science research community~\cite{irrgang2021towards,chen2024collaboration,bracco2025machine}. 
Discriminative models, such as Graph Convolutional Networks and Transformers, can automatically extract complex spatiotemporal dependencies from large scale observational and reanalysis datasets, significantly improving the accuracy and efficiency of deterministic predictions~\cite{kurth2023fourcastnet,lam2023learning}. 
At the same time, generative AI approaches, including diffusion models and generative adversarial networks, have begun to address fundamental limitations of traditional data assimilation frameworks~\cite{geer2021learning,huang2024diffda,brajard2020combining,sonnewald2021bridging,martin2025generative}. 
These models can learn the posterior distribution of meteorological states, enabling the reconstruction of complete physical fields from sparse or noisy observations and supporting next-generation probabilistic forecasting systems. 
Such advances collectively enhance the capacity to dynamically assimilate sparse and heterogeneous observations, improving data completeness, resolution, and spatiotemporal consistency, which are key prerequisites for robust tipping point analysis.
\par

AI holds significant promise for advancing the detection and prediction of tipping points in complex systems.  
Its advantages are reflected in three main aspects.  
First, deep learning models can substantially improve both the sensitivity and specificity of early-warning indicators, while maintaining strong generalizability across diverse systems~\cite{BuryDeep2021}.  
Second, AI methods are particularly well-suited for capturing rate-dependent tipping events that are often missed by conventional EWS approaches, thereby enabling the detection of nonlinear abrupt transitions~\cite{huang2024deep}.  
Third, recent advances in deep learning architectures allow for predictive modeling that extends beyond traditional early warnings, enabling not only the estimation of tipping likelihoods but also the inference of critical thresholds~\cite{liu2024early,zhuge2025deep}.  
Moreover, generative AI techniques can synthesize realistic surrogate data to augment limited observations, thereby enhancing the identification of critical states and improving the robustness and reliability of early-warning frameworks~\cite{ma2025predicting}.  
In addition, the rise of explainable AI~\cite{montavon2019layer} provides a crucial pathway to interpret the inner workings of these otherwise ``black-box'' models,  which offers deeper insights into the mechanisms driving tipping behavior in marine and climate systems.

\subsection{2. Complex networks method}

In parallel, complex network theory has emerged as a powerful tool for analyzing the structure and dynamics of the Earth system~\cite{fan2021statistical}. 
For example, Jiang \emph{et al.}~\cite{jiang2018predicting} constructed complex networks from empirical data and reduced them to equivalent low-dimensional dynamical systems, enabling accurate prediction of tipping events. 
Climate networks, in particular, have proven effective for identifying and characterizing tipping elements. 
Fan \emph{et al.}~\cite{fan2017network} demonstrated that a climate network constructed from global near-surface temperature data could capture localized impacts of El~Niño events, while Liu \emph{et al.}~\cite{liu2023teleconnections} revealed teleconnections between the Amazon rainforest and the West Antarctic Ice Sheet, suggesting that the Tibetan Plateau may represent a potential tipping element. 
Furthermore, network-based indicators, such as normalized degree, average path length, and betweenness centrality, have been applied to detect global tipping behavior~\cite{moinat2024tipping}. 
In addition,  some network models, such as interdependent networks~\cite{BuldyrevCatastrophic2010}, can validly describe the cascading effects in real systems.
Together, these advances underscore that integrating AI methodologies with complex network analysis provides a powerful and data-rich framework for detecting, predicting, and ultimately mitigating tipping and cascading phenomena in the Earth system.
\par

\bibliographystyle{nsr}

\end{document}